\documentclass[useAMS,usenatbib]{mn2e}
\usepackage{graphicx}
\usepackage{natbib}
\usepackage{color}
\newcommand{\feh}{\mbox{[Fe/H]}}
\newcommand{\mgh}{\mbox{[Mg/H]}}

\title[Relationship between metallicity distributions of globular clusters and of circumgalactic gas]
{On the relationship between metallicity distributions of globular
clusters and of circumgalactic gas}
\author[I.A. Acharova and M.E. Sharina]{I.A. Acharova$^{1}$\thanks{E-mail:
iaacharova@sfedu.ru (AIA); sme@sao.ru (SME)} and M.E. Sharina$^{2}$\footnotemark[1]\thanks{}\\
$^{1}$Faculty of Physics, Southern Federal University, 5 Zorge, 344090 Rostov-on-Don, Russia\\
$^{2}$Special Astrophysical Observatory, Russian Academy of Sciences, N. Arkhyz, KCh R 369167, Russia}
\begin{document}

\date{Accepted  Received }

\pagerange{\pageref{firstpage}--\pageref{lastpage}} \pubyear{2002}

\maketitle

\label{firstpage}

\begin{abstract}
The abundance of alpha elements and iron in stars of globular clusters (GCs) shows the composition
of the gaseous medium, in which they have been formed. In the present paper, we discuss a possibility
to consider dense clouds of circumgalactic gas (partial Lyman limit systems  and 
Lyman limit systems) observed in the 100 -- 130~kpc neighbourhood of galaxies at redshifts 
of $\rm0.1<z<1.1$ as being the residual parts of clouds, in which globular clusters have been formed. 
Conclusions have been drawn based on statistical analysis of the abundance of magnesium and iron in GCs 
and in circumgalactic clouds and on the spatial location of objects of both types.
\end{abstract}


\section[]{Introduction}

Interpretation of conditions in which globular clusters (GCs) of ours and other galaxies were formed is a field of 
intense studies by present-day astrophysics, in which there are still many unsolved issues. Over the past decade, 
significant progress has been made in understanding the properties of GCs and the conditions in which they were 
formed (see the review by \citet{k14}). 

It has long been established in the number of studies that GCs of the Galaxy are divided into two distinct subsystems: 
metal-rich and metal-poor in  terms of iron \citep{h96, d16, c10} and magnesium abundances \citep{d16, c10, p05}. 
This result confirms the earlier result obtained on the basis of the study of metallicity 
\citep{ms77, z85}. The data, suitable for statistical analysis of the abundance of iron and magnesium 
in GCs, allow us to make significant progress in the study of early stages of evolution of galaxies. 
It was argued that these elements are produced by two different types of supernovae mainly \citep{ww95, t95, lc03,h05}: 
magnesium is formed as a result of explosions of core-collapse supernovae (\mbox{SNe~CC}), iron is formed by type I a supernovae (\mbox{SNe~Ia}).

 The average ages of GCs in the metal-rich and the metal-poor groups are similar \citep{mf09, vb13, c10, c12} within measurement errors 
which can reach 25\%. According to \citet{c12}, the mean age of the metal-rich Galactic and M31 GCs is $10.6 \pm 0.5$ Gyr, 
and the mean age of the metal-poor Galactic and M31 GCs is $10.2 \pm 0.2$ Gyr.
It follows therefrom that the enrichment of gas with magnesium and iron occurred quickly. 
There is nothing surprising 
in this rapid enrichment by magnesium: the lifetime of \mbox{SNe~CC}  progenitors from birth to explosion is only a few million years. 
Rapid iron enrichment  was difficult to explain up to and including 2005, because it is thought that the time interval from 
the moment of formation and up to the explosion in \mbox{SNe~Ia} progenitors is about 1.5 billion years, and they massively have exploded 
in 3 billion years \citep{d04, s05}. It has been established in recent years that the age of \mbox{SNe~Ia} 
progenitors is  $T<500$~Myr and the frequency of \mbox{SNe~Ia} explosions in star-forming regions is 3-5 times higher than the corresponding 
frequency in  the representatives of old stellar populations  \citep{m06,b10, l11, m14, r14, ah18}.

The present study discusses a possibility to consider circumgalactic clouds (partial Lyman limit systems, pLLSs, 
with neutral hydrogen column densities in the range $16.1<logN_{HI}<17.2$ and  Lyman limit systems, LLSs, $17.2<logN_{HI}<17.7$) 
observed in the 100--130~kpc vicinity from galaxies at redshifts of $\rm0.1<z<1.1$ as the relic of the gaseous medium, 
from which GCs were formed.
  LLSs and pLLSs are highly ionised gaseous structures in distinction to more dense Damped Ly$\alpha$ systems (DLAs; $logN_{H I} > 20.3$)
which are partly neutral. LLSs and pLLSs are expected to probe cool,
dense streams through the circumgalactic medium of galaxies (\citep{w16} and references therein). 
In 2013, the results of the first large-scale studies of the chemical element abundance 
in clouds that appeared in the line of sight of quasars were obtained \citep{l13}. These studies were continued in 
a series of papers \citep{w16, l16}. It has been found that the magnesium abundance distribution estimated 
from 55 clouds is bimodal, i.e., the distribution function has two distinct peaks with the deep minimum at $\rm\mgh \sim -0.9$. 
The statistics of these studies allows us to draw important conclusions which will be discussed in the paper.
\begin{figure*}
\includegraphics[scale=1.1]{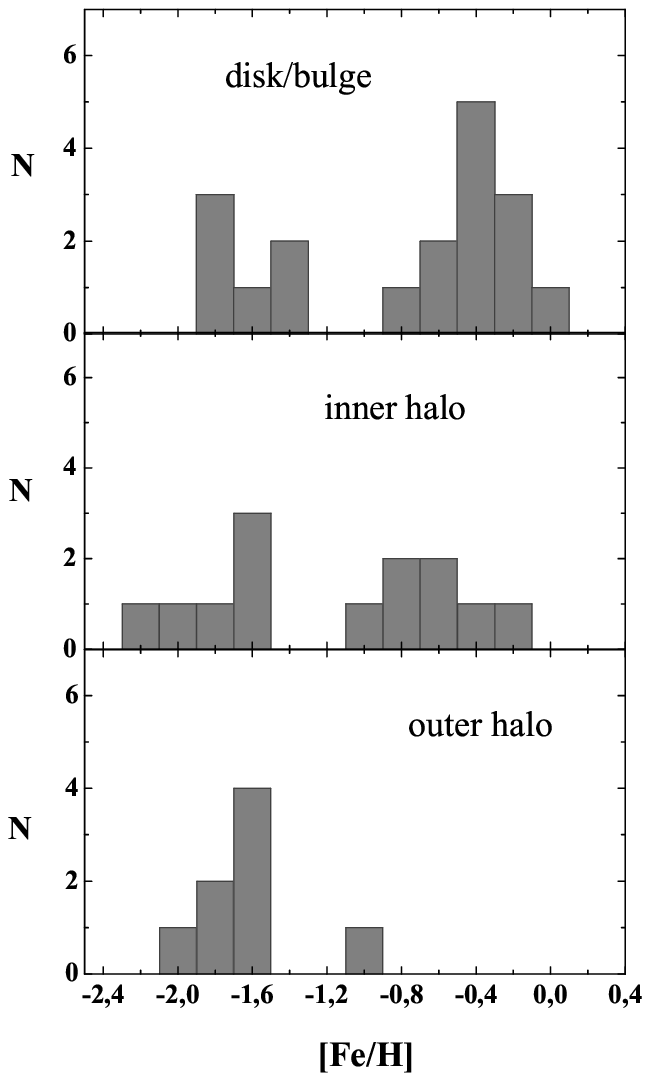}
\includegraphics[scale=1.1]{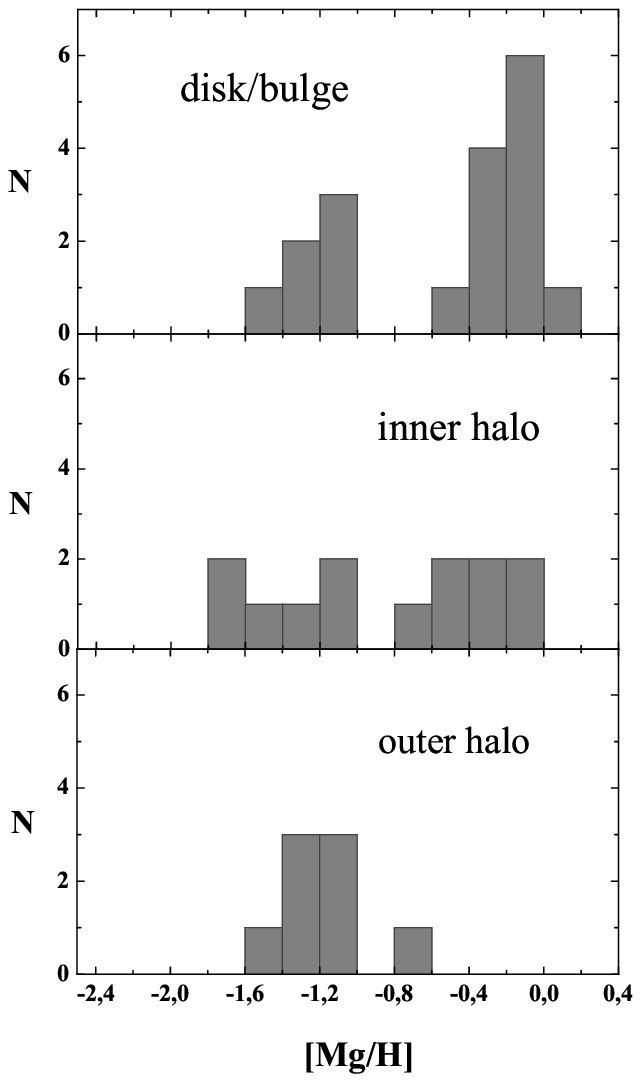}
\caption{Histograms of iron (left-hand) and magnesium (right-hand)abundances in GCs built using the data from  \citet{d16} for objects 
of three subsystems of the Galaxy: disk/bulge objects, of the inner and outer halos.}
\label{fig1}
\end{figure*}

\section[]{Statistical analysis of iron and magnesium abundances in globular clusters}

Our conclusions about the abundance of chemical elements in GCs are based on analysing the results 
of the studies presented in three papers: \citet{d16, c10, p05}.
The paper by \citet{d16} presents a sample of fifty one Galactic GCs with parameters determined using one method. 
The abundances of two chemical elements of interest, \feh\ and \mgh, were determined using the MILES library. 
The authors divided clusters into three groups according to the criteria proposed in \citet{c10}: objects of 
the disc/bulge and of the inner and outer halos. As can be seen from the analysis of histograms built separately 
for  disc/bulge and inner halo objects (see Fig.~1), the distributions are bimodal with a deep minimum at 
$\feh \sim-1.2$ ($\mgh \sim-0.9$). Tables 1 and 2 give average values of the iron and magnesium abundance, 
root-mean-square deviations, and the number of GCs in each group.  
Let us note that iron-poor clusters coincide with those being magnesium-poor.

Let us proceed to the study carried out by \citet{c10}. In that paper, 139 GCs from the Harris catalogue \citep{h96} were 
divided into three groups on the basis of the cluster velocity and their position in the Galaxy. 
To the outer halo, twenty eight GCs were assigned, to the inner halo -- thirty five GCs, to the disc with a bulge -- seventy six GCs. 
The histograms of the iron abundance in the GCs of three groups are shown in Fig.~2. It can be noticed that two 
sub-populations are clearly seen only in the disc/bulge group: metal-poor consisting of thirty nine GCs with the average value 
$\rm\langle\feh\rangle=-1.50$ and metal-rich consisting of thirty seven GCs with the average value  $\rm\langle\feh\rangle=-0.44$. 
Between two cluster groups, there is a deep minimum in the vicinity of $\rm\feh \sim-1.2$. Objects of the outer and inner halos have 
a number of measurements sufficient for statistical analysis in the metal-poor region only. Mean values of the iron abundance, 
the root-mean-square deviations, and the number of GCs in each group are given in Tables 1 and 2.

In the paper by \citet{c10}, $\rm\mgh$ were additionally determined for nineteen GCs. Fifteen of them are metal-poor with $\rm\feh<-0.9$. 
This group of clusters includes objects of both the halo and disc subsystems. Taking into consideration a small number of GCs, 
we did not divide them into subgroups and built histograms for the entire sample (Fig.~3). The average values and standard 
deviations in the metal-poor and metal-rich regions are given in Tables 1 and 2.

\begin{figure}
\includegraphics[scale=1.2]{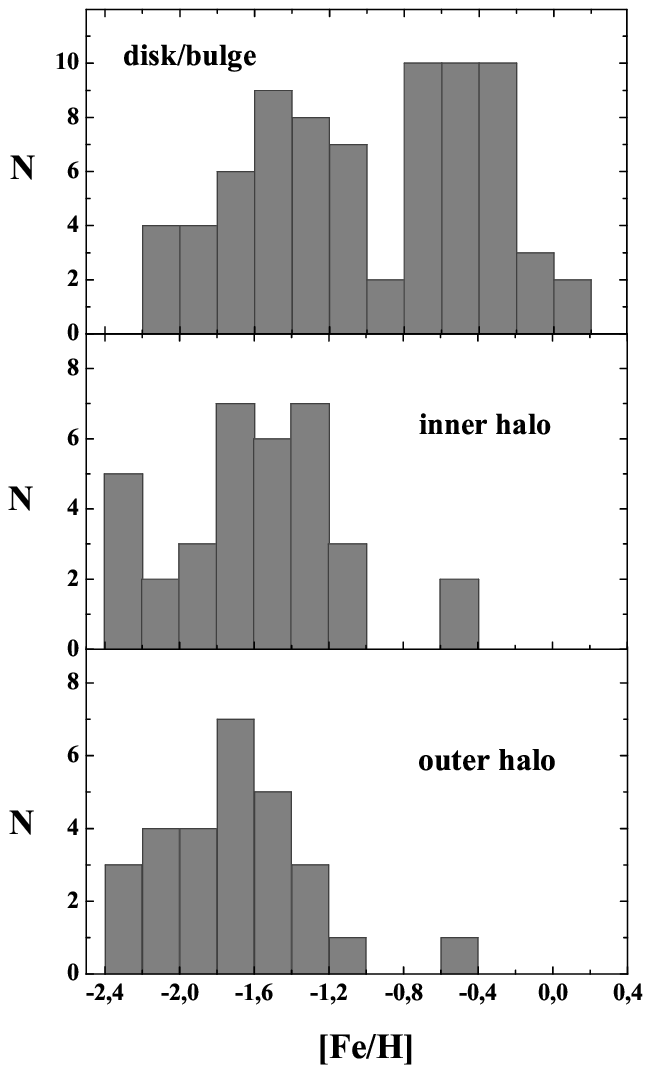}
\caption{Distribution of the iron abundance in GCs built from the \citet{c10} data for the objects of three subgroups of GCs in our Galaxy: 
objects of the disc/bulge and of the inner and outer halos.}
\label{fig2}
\end{figure}
\begin{table*}
 \centering
 \begin{minipage}{140mm}
\caption{Iron abundance in  globular clusters.}
\begin{tabular}{@{}l|ccc|ccc@{}}
\hline \hline
 subsystem  &  \multicolumn{3}{c}{metal-poor group} & \multicolumn{3}{c}{metal-rich group} \\ %
 in the Galaxy &           &      &                 &         &        &           \\ \hline
                 & $ <\feh> $ & $ \sigma(\feh)$ & number of GCs& $ <\feh> $ & $ \sigma(\feh)$ & number of GCs \\ \hline
\multicolumn{7}{c}{Carretta et al. 2010} \\ \hline
19 GCs with & -1.69 & 0.41 & 14 & -0.62 & 0.17 & 5 \\
measured \mgh&           &      &                 &         &        &           \\
inner halo                 & -1.67 & 0.37 & 33 & -- & -- & 2 \\
outer halo &-1.74 & 0.32 & 27 & -- & -- & 1 \\
disc/bulge & -1.50 & 0.33 & 39 & -0.44 & 0.26 & 37 \\ \hline
\multicolumn{7}{c}{Dias et al. 2016}\\ \hline
inner halo & -1.78 & 0.24 & 6 & -0.64 & 0.24 & 7 \\
outer halo  & -1.70 & 0.13 & 7 & -- & -- & 1 \\
disc/bulge & -1.65 & 0.14 & 6 & -0.38 & 0.21 & 12 \\ \hline
\multicolumn{7}{c}{Pritzl et al. 2005}\\ \hline
halo       & -1.73 & 0.37 & 32 & -0.54 & 0.20 & 3 \\
disc/bulge & - & - & - & -0.55 & 0.29 & 6 \\ \hline
\hline
\end{tabular}
\end{minipage}
\label{tab1}
\end{table*}

\begin{table*}
 \centering
 \begin{minipage}{140mm}
\caption{Magnesium abundance in globular clusters and circumgalactic clouds.}
\begin{tabular}{@{}l|ccc|ccc@{}}
\hline \hline
 subsystem  &  \multicolumn{3}{c}{metal-poor group} & \multicolumn{3}{c}{metal-rich group} \\ %
 in the Galaxy &           &      &                 &         &        &           \\ \hline
               & $ <\mgh> $ & $ \sigma(\mgh)$ & number of GCs& $ <\mgh> $ & $ \sigma(\mgh)$ & number of GCs \\ \hline
  \multicolumn{7}{c}{Carretta et al. 2010} \\ \hline
  19 GCs with & -1.19 & 0.35 & 15 & -0.17 & 0.15 & 4 \\
measured \mgh&            &      &                 &         &        &           \\ \hline 
\multicolumn{7}{c}{Dias et al. 2016}\\ \hline
inner halo & -1.37 & 0.25 & 6 & -0.37 & 0.17 & 7 \\
outer halo  & -1.27 & 0.13 & 7 & -- & -- & 1 \\
disc/bulge & -1.21 & 0.15 & 6 & -0.19 & 0.14 & 12 \\ \hline
\multicolumn{7}{c}{Pritzl et al. 2005}\\ \hline
halo  & -1.38 & 0.42 & 33 & -- & -- & 2 \\
disc/bulge & -- & -- & -- & -0.24 & 0.23 & 6 \\ \hline
\multicolumn{7}{c}{Wotta et al. 2016}\\ \hline
pLLSs & -1.58 & 0.23 & 24 & -0.34 & 0.28 & 20 \\
pLLSs$+$ &-1.49& 0.28 & 32 & -0.38 & 0.30 & 22 \\
LLSs &            &      &                 &         &        &           \\  \hline
\hline
\end{tabular}
\end{minipage}
\label{tab2}
\end{table*}
Let us turn to another study performed in \citet{p05}. It shows determinations of the abundances of chemical elements in forty five GCs 
obtained by several groups of authors based on high-resolution spectroscopy of red giants (see Tables 1 and 2 in their paper). 
For most GCs, the abundance of a chemical element in question was performed by different authors. We used the average values in 
their Table 2. For our purpose, we need GCs with the iron and magnesium abundances measured. Such data are available for forty one GCs. 
The clusters belong to different galactic subsystems. According to the criteria proposed in the paper by \citet{mg04}, 
nineteen GCs are classified as belonging to the disc/bulge subsystem including the thick disc, the rest -- to the halo subsystem. 
In compliance with the studies of the cluster-velocity components  \citep{p05}, six GCs are  referred to the disc/bulge subsystem, 
the rest of objects were divided into the young and old halo subsystems, including 3 objects of the Sagittarius dSph galaxy (M54, Ter7, Pal12). 
Comparing both classifications, one can notice that according to \citet{p05} the halo objects are referred to thick-disc objects, 
while according to \citet{mg04} all them belong to the metal-poor subgroup with the exception of three GC: Ter7, Pal 12, and NGC6553. 
For our purpose, it is necessary to determine the average abundance of iron and magnesium for the metal-poor and 
metal-rich components of a GC, thus, we confine ourselves to analysing the properties of GCs divided into groups in compliance 
with the classification presented in \citet{p05}.  Tables 1 and 2 show the average abundance, root-mean-square deviations, 
and the number of clusters in the metal-poor and metal-rich subgroups.

The bimodal metallicity distribution of GCs is also observed in other massive galaxies. It is known, for example, that in the 
Virgo Cluster galaxies the metallicity distribution of GCs is also bimodal with the average metallicity slightly depending on 
the parent galaxy mass and close to the GC metallicity distribution in our Galaxy \citep{k14}. The metallicity 
distribution of GCs in M31 and of ultracompact dwarf galaxies also complies with this conclusion (\cite{c12} and references therein).

The average iron and magnesium abundances of metal-poor GCs determined by the results of all the authors, weighted according 
to the number of objects, are $\rm\langle\feh\rangle=-1.66\pm 0.34 $, $\rm\langle\mgh\rangle=-1.31\pm 0.35$. The same for metal-rich GCs: 
$\rm\langle\feh\rangle=-0.47 \pm 0.25$, $\rm\langle\mgh\rangle=-0.24\pm 0.18$.

\begin{figure*}
\includegraphics[scale=0.8]{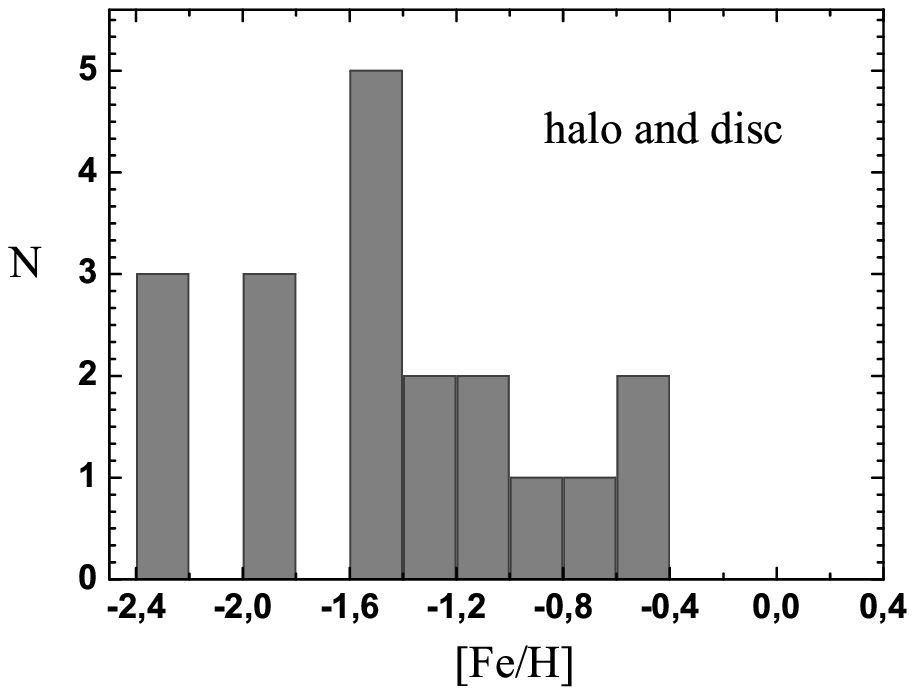}
\includegraphics[scale=0.8]{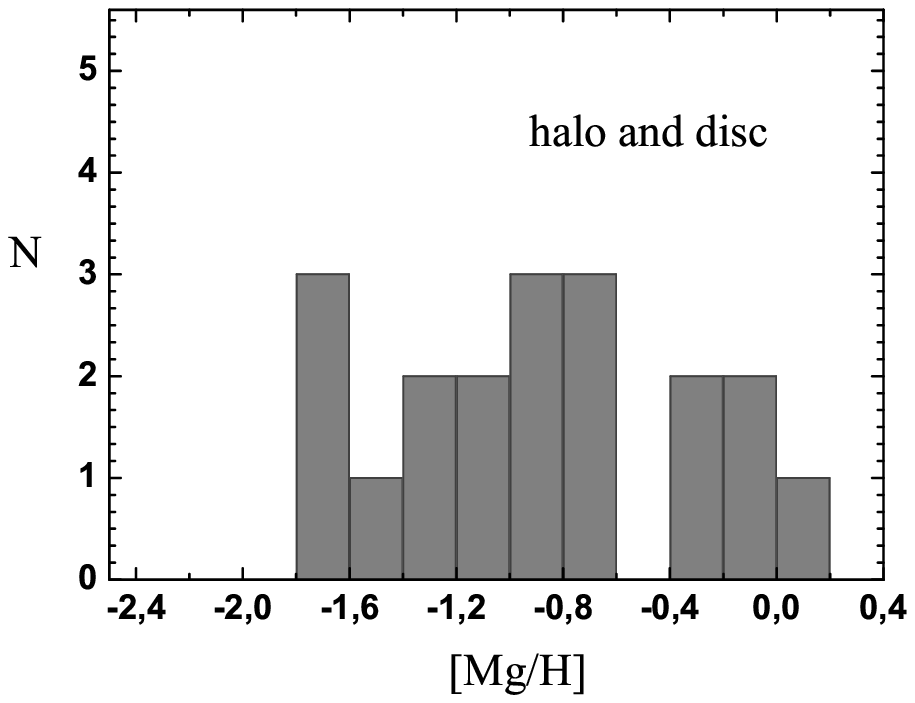}
\caption{Distributions of iron (left-hand) and magnesium (right-hand) abundances for nineteen GCs of both the halo and disk subsystems from \citet{c10}, 
for which magnesium abundances are measured.}
\label{fig3}
\end{figure*}

\begin{figure*}
\includegraphics[scale=0.9]{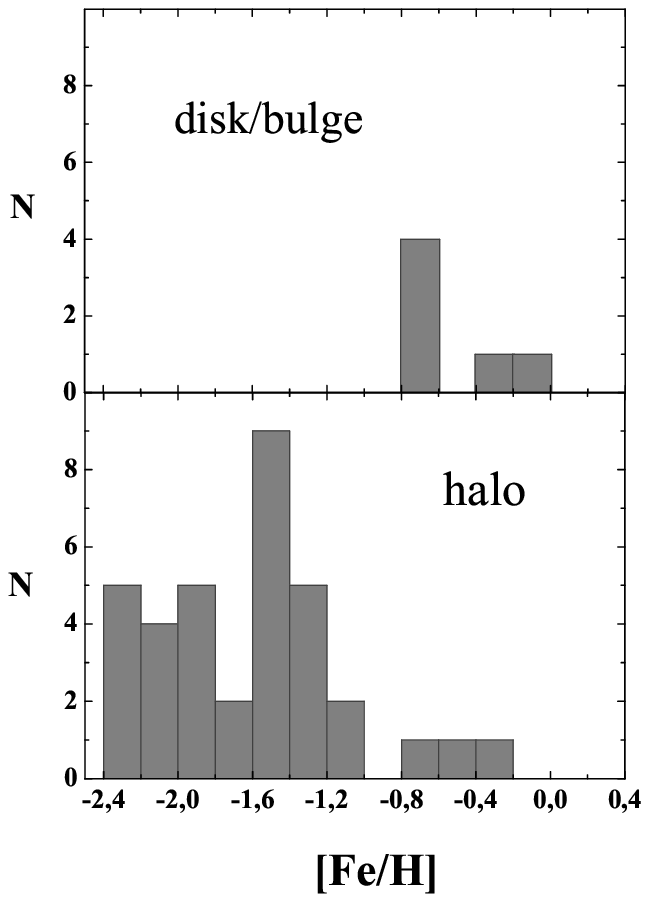}
\includegraphics[scale=0.9]{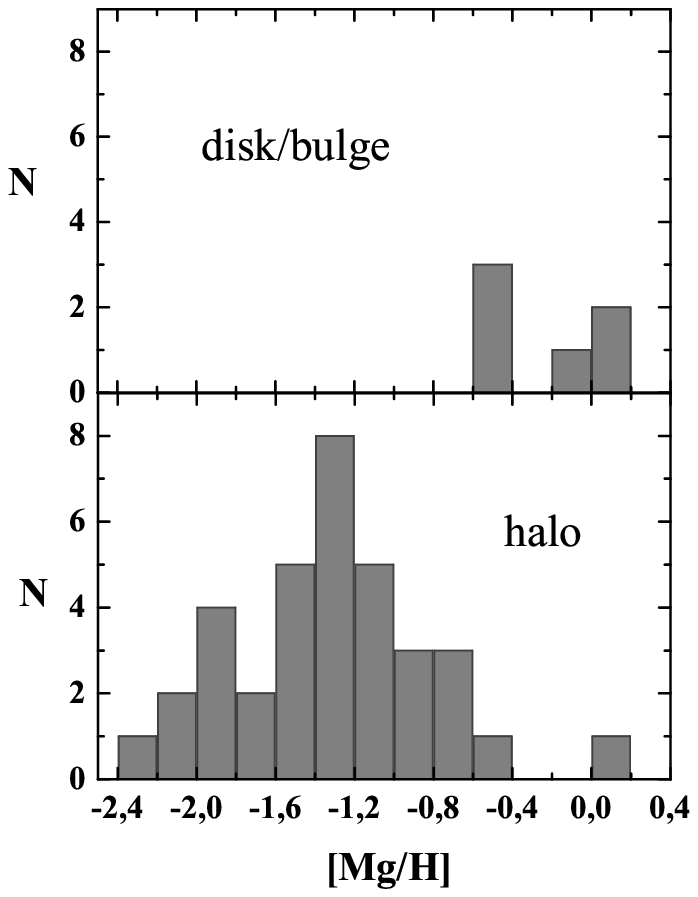}
\caption{Distributions of iron (left-hand) and magnesium (right-hand) abundances in GCs built using the data from \citet{p05} for objects 
in the two subsystems of the Galaxy: halo and disc/bulge objects.}
\label{fig4}
\end{figure*}

\section[]{Statistical analysis of magnesium abundance in clouds of the circumgalactic medium}

The bimodal metallicity distribution of clouds of the circumgalactic medium was for the first time established 
by \citet{l13}. The first results of analysing the metallicity of the circumgalactic medium of 
galaxies located at redshifts of $\rm z<1$ were presented and it was obtained that the metallicity distribution 
estimated from the alpha-element abundance in twenty eight Lyman limit systems and partial Lyman limit systems is bimodal, 
i.e., the distribution function has two distinct peaks with the deep minimum at $\rm\mgh \sim-0.9$. The bimodal metallicity 
distribution in galaxies at redshifts from one to zero means that the metal-poor and metal-rich gas did not noticeably mix 
during this time interval. All the clouds are located within 100~kpc from the centers of the nearest galaxies.

The paper by \citet{w16} continued these studies. Its authors analysed the magnesium abundance in the dense 
circumgalactic medium for galaxies with redshifts of $\rm0.1<z<~1.1$. The lines of singly-ionized magnesium are strong 
and well-resolved, thus, this element is an exceptional indicator of metallicity, in the authors opinion. 
Combined with the studies of 2013, forty four optically thin systems (pLLSs) and eleven optically thicker (LLSs) have been investigated. 
It has been found that the metallicity distribution of pLLSs at $z<~1$ is bimodal, therefore, the distribution function 
has two distinct peaks, each of which is close to the normal distribution with a dip between them at $\rm\mgh \sim-0.9$. 
In the sample observed, the maximum of the distribution function for the metal-poor group of pLLSs is at $\rm\mgh \sim-1.58$, 
for the metal-rich group it is at $\rm\mgh \sim-0.34$ (see Fig.~5, the analogue of Fig.~12 in \citet{w16}).

Fig.~6 shows the metallicity distribution of circumgalactic clouds at redshifts of $z<1$ for the whole sample pLLSs+LLSs ($16.1<logN_{HI}<17.7$). 
The metallicity distribution function is bimodal with the distinct minimum at $\mgh \sim -0.9$. Table 2 gives the number of clouds 
in each subgroup, the average abundance of magnesium, and the root-mean-square deviations.

The average abundance of magnesium in the sample of clouds consisting of pLLSs$+$LLSs in the metal-poor group is $\rm\mgh=-1.49$, 
in the metal-rich group -- $\rm\mgh=-0.38$. Comparing the magnesium abundance in the metal-poor and metal-rich subgroups of circumgalactic 
clouds and the magnesium abundance in the same subgroups of GCs, one can conclude that they coincide within errors of abundance 
measurements in clouds of 0.3 dex. For the metal-rich group, the magnesium abundance in clouds is by 0.14 dex smaller than in GCs. 
For the metal-poor subgroup, the magnesium abundance in clouds is by 0.18 dex smaller than in GCs. Hence, one can assume that  
pLLSs+LLSs clouds can be the rest of parent clouds of GCs. 
It is shown in the paper by \citet{l16} that
 high-metallicity pLLSs$+$LLSs appear starting from $z\sim2.5$ that corresponds to the time of their formation 
$T\sim11$ Gyr ago. 
At redshifts of $2.5<z<3.3$, the metallicity distribution for clouds is 
presented mainly by a metal-poor component. It follows herefrom that, heavier-element enrichment of the fraction of metal-poor gas has 
taken place through thermonuclear fusion products of supernovae of the first GC generation entering into it. 
The second generation has subsequently been formed from this enriched gas.
 It should be noted that the ages of some high-metallicity Galactic GCs were estimated to be 12-13~Gyr (see Table 1 in \citet{c12}), 
 i.e., they are as old as low-metallicity GCs according to these data.
The errors of absolute age determination are large and can reach 25\%. It is not excluded that the process of star cluster formation
was extended in time, and low-metallicity GCs continued to form at the same time, when high-metallicity GCs 
originated in other clouds. The picture is complicated by the fact that the expanding shells from
supernovae that have exploded in low-metallicity GCs are fragmented due to
enhanced gaseous instabilities 
\citep{a02, dsh04, vds09} and radiation losses \citep{v17}
in high-metallicity and low-metallicity regions. The characteristic time of metal mixing between such fragments is hundreds 
millions of years  \citep{a02}, which is much greater than the typical time-scale of a star-forming burst which is about 
several million years. Therefore, it is impossible to completely exclude the formation of low-metallicity GCs 
at the second stage of star formation in the clouds. 
It is difficult to reconstruct the exact evolutionary scenario due to age determination uncertainties.

\section[]{Estimation of magnesium production by supernovae in the first generation of globular clusters and enrichment of the parent cloud with metals}

\begin{figure}
\includegraphics[scale=0.85]{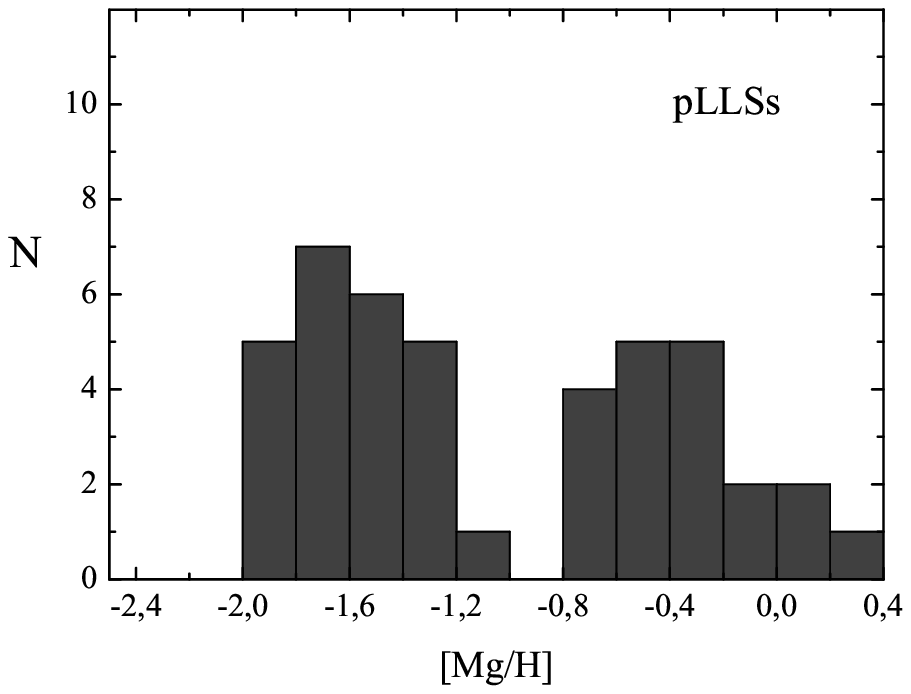}
\caption{Magnesium abundance distribution in 44 partial Lyman limit systems at redshifts of $0.1<~z<~1.1$ determined in \citet{w16}. 
The metallicity distribution is bimodal. The distinct dip in the distribution is observed at \mgh=-0.9.}
\label{fig5}
\end{figure}
\begin{figure}
\includegraphics[scale=0.85]{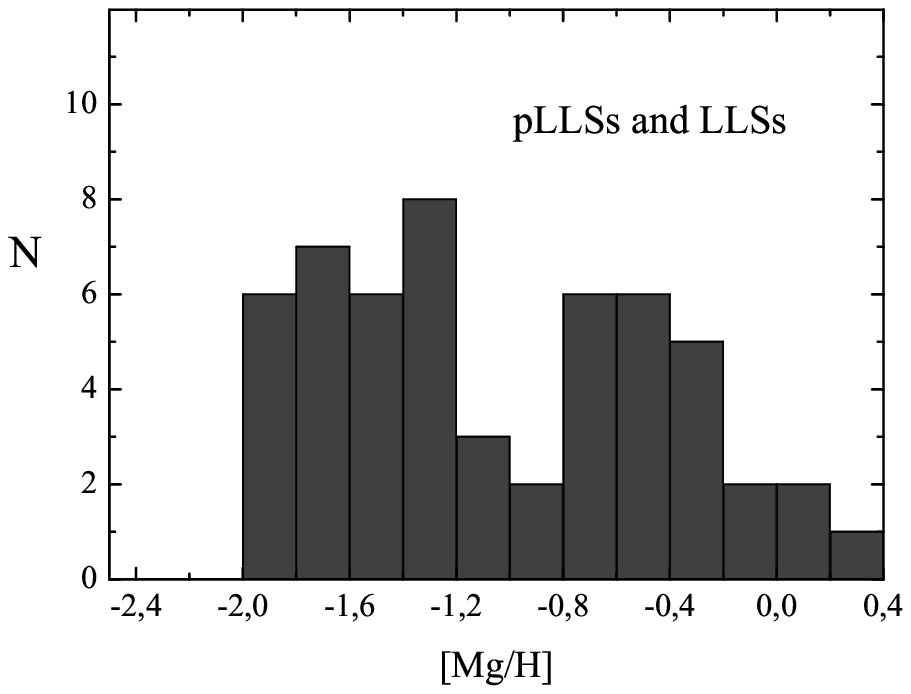}
\caption{Magnesium abundance distribution in 55 partial Lyman limit systems and Lyman limit systems at redshifts of $0.1<~z<~1.1$ determined in \citet{w16}. 
The metallicity distribution is bimodal. The distinct dip in the distribution is observed at \mgh=-0.9.}
\label{fig6}
\end{figure}
Let us estimate the mass fraction of magnesium corresponding to the average values of  \mgh\ for metal-rich and metal-poor groups of 
GCs and circumgalactic clouds.

Let $Z_{Mg}$ denote the mass fraction of magnesium in relation to hydrogen in the object under study,
$Z_{Mg_{\sun}}$ is the same mass fraction in the Sun. $\rm\mgh$ can mean both the fraction by the number of atoms $N_{Mg}\over N_{H}$
and by the mass, where the mass fraction is determined with the expression $\rm N_{Mg} \cdot m_{Mg}\over N_{H} \cdot m_{H}$, 
with the ratio of masses of magnesium and hydrogen atoms decreasing with the logarithm difference: $\rm \mgh = lg Z_{Mg} - lg Z_{Mg_{\sun}} $.
It follows from here that $Z_{Mg} = 10^{lg Z_{Mg_{\sun}} + \mgh}$.

The mass fraction of magnesium in the Sun $Z_{Mg_{\sun}} = {{N_{Mg} \cdot m_{Mg}}\over {{N_{H} \cdot m_{H}}}} X$,
 where $X=0.7381$ is the mass fraction of hydrogen in the Sun \citep{a09}. It follows from here that 
 $lg Z_{Mg_{\sun}} = lg ({N_{Mg}\over N_{H}}) + lg ({m_{Mg}\over m_{H}}) + lg(X) = -3.1$.
 (We used the value determined in the paper by \citet{a09} for the Sun 
 $A_{Mg} = lg ({N_{Mg}\over N_{H}}) + 12 = 7.64$ ).
Thus, if the magnesium abundance in GCs is $\mgh=-1.31$, then the mass fraction of magnesium in it is $Z_{Mg}=10^{-1.31-3.1} = 10^{-4.41}$. 
If  the magnesium abundance in GCs is $\mgh=-0.24$, then the mass fraction of magnesium in it is $Z_{Mg}=10^{-0.24-3.1} = 10^{-3.34}$
Therefore, the magnesium abundance in the metal-rich GC group is twelve times higher than that in the metal-poor group.

Similar reasoning is correct for circumgalactic gaseous clouds also. The mass fraction of magnesium in the metal-rich group equals
$Z_{Mg}=10^{-0.38-3.1} = 10^{-3.48}$, in the metal-poor group $Z_{Mg}=10^{-1.49-3.1} = 10^{-4.59}$.
In the metal-rich group of circumgalactic clouds, the magnesium abundance is approximately thirteen times higher than that in the metal-poor one.

 Thus, taking into consideration the similarity of metallicity distributions of GCs and circumgalactic clouds for $z<1$ and the prevalence 
of the metal-poor component for $z>2$, we can suppose that the metal-rich component originated from the metal-poor one through 
formation of GCs with subsequent enrichment with supernovae explosion products.

\subsection{Limitations imposed by GCs on magnesium and iron nucleosynthesis}
\label{sec_4_1}
In the papers by \citet{k14} and \citet{s10}, it is shown that the regions of gas concentration observed 
at high redshifts around large galaxies with the characteristic sizes of 1-3 kpc and masses $M \propto 10^8 - 10^9 M_{\sun}$
can be the birthplaces of several hundreds of GCs, only a dozen of which have survived till our days. 
For their part, \citet{l13}, having qualitatively estimated the properties of circumgalactic clouds, drew the conclusion that 
their sizes could reach several kpc.

 As was shown in the paper by \citet{shch00}, the process of gas enrichment with supernova explosions occurs in two stages.
Supernova bursts push chemical elements out of GCs or UCDs. The expanding shells lose energy due to the high pressure of the 
surrounding gas and form bubbles of a relatively small size. Modelling of shell expansion is illustrated, for example, in 
\citet{shen14} and \citet{v17}. 
At the second stage, metals are diffused over long distances
comparable to the distance between galaxies.
As it was shown by \citet{shch00}, turbulent mixing of gaseous layers at the interface between shock 
waves from supernova can be important for determining the spatial structure of the metal distribution.
The areas enriched with metals gradually increase in number and size. Unfortunately, 
the details of this process are not yet clear.
Thus, now we observe the relics of the primeval clouds which served as the building material for metal-poor GCs, 
and the remains of the metal-enriched parts of these clouds, from which the high-metallicity GC group was formed.

It was shown by \citet{l13} that for the metal-poor sample of pLLSs and LLSs the  number of hydrogen atoms along 
the line of sight $logN_{H}=19.0 \pm 0.5$, while for the metal-rich
sample $logN_{H}=18.6 \pm 0.7$. The sizes and covering factors are similar.
Therefore, the metal-enriched gas in the circumgalactic medium
may be on average less massive than the metal-poor gas by a factor of $\sim 2$ \citep{l13}.

Let us estimate the number of supernovae capable to enrich half the mass of the primeval cloud up to 
the observed abundances of chemical elements.
If the mass of a cloud $ M \approx 10^9 M_{\sun}$, then the magnesium mass in the metal-poor cloud 
$M_{Mg}=10^{9-3.1-1.49} = 10^{4.41} M_{\sun}$. Taking into account that the mass of the metal-rich
cloud is twice smaller $ M \approx 10^{8.7} M_{\sun}$, the mass of Mg in it is
 $M_{Mg}=10^{8.7-3.1-0.38} = 10^{5.22} M_{\sun}$. 
If we assume that a metal-rich cloud with $\mgh=-0.38$ has been formed from a metal-poor one with $\mgh=-1.49$,
then the cloud has acquired $M_{Mg}\approx 10^{5.2} M_{\sun}$.

It is natural to expect that the major contribution in magnesium production has been made by rapidly evolving massive stars
that exploded as \mbox{SNe~CC} . According to calculations of \citet{t95} and \citet{h05}, a single core-collapse supernova  ejects
$M_{Mg} \approx 0.1 M_{\sun}$ when exploding. In order to produce $M_{Mg} \approx  10^{5.2} M_{\sun}$, $ 10^{6.2}$ \mbox{SNe~CC} 
are necessary. For the reasoning given, the accepted mass of a cloud is not critical, since both the magnesium mass in the cloud 
and the number of supernovae in it are proportional to the mass of the cloud: the cloud with $M \approx 10^8 M_{\sun}$
will obtain the magnesium mass $M_{Mg} \approx 10^{4.2} M_{\sun}$ with $10^{5.2}$ \mbox{SNe~CC} required.

Let us estimate whether it is possible to expect such a number of \mbox{SNe~CC} resulting from star formation. As is shown in the paper by 
\citet{s10} (and references therein), structures with a mass of $10^{-6} - 10^{-3}$ of the gaseous clouds mass have been formed 
inside it at the initial stages of evolution of a galaxy; and, what is actually important for our study,
about 10\% of the mass of a cloud is transferred to stars \citep{w03}, from which the first generation of GCs can later develop. 
Namely, the mass of stars formed from a cloud with $M \propto 10^9 M_{\sun}$ will be $M_* \propto 10^8 M_{\sun}$.

The stellar mass distribution is described by the initial mass function (IMF). The IMF is conventionally denoted as $\varphi(m)$
and determined in such a way that the mass fraction of stars enclosed in the interval [$m; m+dm$] 
is described with the expression $m \varphi(m) dm$. The integral $\!\!\int_{m_{l}}^{m_{u}}\!\! m \varphi(m) dm =1$ is a condition 
of IMF normalizing, where $m_l$, the minimum mass of forming stars, is assumed to be equal to $0.1 M_{\sun}$, $m_u$ --
the maximum mass of forming stars is assumed to be equal to $70 M_{\sun}$.

We are interested in stars that can explode as \mbox{SNe~CC}. Such stars have an initial mass of greater than $8 M_{\sun}$.

It was found in recent papers \citep{w15} (and references therein) 
that the high-mass IMF slope ($m>1.5 M_{\sun}$) inferred from numerous
studies of young star clusters is in the range of 1.2$\div$1.7. 
The IMF slope is constant and equal to 1.35 in case if we choose the Salpeter IMF \citep{sal55}.  
\citet{w15} concluded that for 85 star clusters
in M31 the high-mass slope is universal and equal to $1.45 ^{+0.03}_{-0.06}$. 
It is difficult to use the findings of \citet{w15, kr01, kr03} and  \citet{zhang99} in our calculations, 
because the normalization coefficients are not given. For
the multi-slope expression of IMF, the normalization coefficients may be different for various mass ranges, 
as it is shown in \citet{k93} and \citet{sc86}.
Therefore, in the following we will use the IMF shape proposed by \citet{sal55}.
Let us recall that it was defined while studying  young star clusters: 
$\varphi(m)=0.1716 \cdot m^{-2.35}$. Note, however, that differences between the results of our calculations 
using the mentioned various IMF slopes are within 5\%.

Let us estimate the stellar mass fraction that is accounted for by \mbox{SNe~CC}: 
$\!\!\int_{8}^{70}\!\! m \varphi(m) dm \approx 0.12$.
There are studies \citep{a13, k08} showing that the maximum mass of stars resulting 
in \mbox{SNe~CC} can not exceed $30 M_{\sun}$. In this case, only 
$\!\!\int_{8}^{30}\!\! m \varphi(m) dm \approx 0.09$ of the GC mass is accounted for by \mbox{SNe~CC}.  
Thus, about 10\% of GC mass is accounted for by \mbox{SNe~CC} in two considered cases.
Consequently, it can be concluded that the mass of supernovae formed from a cloud with 
$M \propto 10^9 M_{\sun}$ will be $M_* \approx 10^7 M_{\sun}$.

The average mass of a star that explodes as a supernova is found from formula (1) from \citet{t95}:

$\langle Mcc \rangle = {{\!\!\int_{8}^{70}\!\! m \varphi(m) dm}\over{\!\!\int_{8}^{70}\!\! \varphi(m) dm}} \approx 17.5$~$M_{\sun}$
($\langle Mcc \rangle = 15 M_{\sun}$  for $m_u=30 M_{\sun}$).

It follows from here that approximately $10^{5.8}$ \mbox{SNe~CC} can explode in the cloud, which is  2.5 times 
smaller than the required number of supernovae for the thirteen times enrichment of
one half of the cloud with magnesium. Such an amount of \mbox{SNe~CC}  will generate $M_{Mg}\approx  10^{4.8} M_{\sun}$.

Three possibilities to solve the issue of producing the necessary mass of magnesium can be seen. 
The first one is the following: 
 the fraction of gas in the cloud enriched by the products of first supernovae  constitutes not 50\% but 20\% 
 ($M \approx 10^{8.3} M_{\sun}$) of the initial cloud mass.
 This assumption looks believable taking into account the errors of measuring $logN_{H}$ \citep{l13}.
 Then the mass of magnesium in the enriched part of the cloud will be $M_{Mg}=10^{8.3-3.1-0.38} = 10^{4.82} M_{\sun}$.
The second possibility is that the fraction of the cloud gas transformed into stars should be about 25\%.
This is much, at first glance, 
but as is shown in the papers by \citet{ke12} (and references therein) and \citet{w09}, 
it is observationally established that at a surface density of $100 - 300 M_{\sun}/$pc$^2$ 
a so-called explosive star-formation mode occurs, in which the fraction of gas transformed 
into stars can exceed 50\%. Theoretical studies confirm such a possibility \citep{ms76, m89}. 
The third possibility is the production of the residuary amount of magnesium by another type of supernovae, 
\mbox{SNe~Ia}. According to available models, in which \mbox{SNe~Ia} is considered the result of a white dwarf explosion, 
the magnesium mass emitted during the explosion is approximately equal to 
$M_{Mg}\approx 0.01 M_{\sun}$ \citep{t95}. Therefore, to produce the missing 
$M_{Mg}\approx 10^{5} M_{\sun}$, the number of \mbox{SNe~Ia} should be 6 times higher than the number of \mbox{SNe~CC}. 
Before omitting this alternative, let us consider the analysis of the iron abundance in GCs, 
since \mbox{SNe~Ia} are thought to be the main supplier of iron.

\subsection{Analysis of the enrichment of globular clusters with iron}

In a similar way as conducted for magnesium, we will analyse the iron abundance in GCs. 
Since there is no data on the iron abundance in circumgalactic clouds, 
then the conclusions about the iron abundance in GCs are generalized to the iron abundance 
in the gaseous medium from which they have been formed. 
Let us denote $Z_{Fe}$ -- the iron mass fraction in relation to hydrogen in the objects 
under study,  
$Z_{{Fe}_{\sun}}$ is the iron mass fraction in the Sun. It follows that 
$\rm Z_{Fe}= 10^{lg Z_{{Fe}_{\sun}}+[Fe/H]}$.

The mass fraction of iron in the Sun 
$Z_{Fe_{\sun}} = {{N_{Fe} \cdot m_{Fe}}\over {{N_{H} \cdot m_{H}}}} X$, where $X=0.7381$
is the hydrogen mass fraction in the Sun \citep{a09}. Hence, 
$lg Z_{Fe_{\sun}} = lg ({N_{Fe}\over N_{H}}) + lg ({m_{Fe}\over m_{H}}) + lg(X) = -2.84$
(We used the value defined in the paper of \citet{a09} for the Sun 
$A_{Fe} = lg ({N_{Fe}\over N_{H}}) + 12 = 7.54$).
Thus, if the iron abundance in a GC $\feh=-1.66$,  then the iron mass fraction in it 
$Z_{Fe}=10^{-1.66-2.84}=10^{-4.50}$.
If the iron abundance in a GC $\feh=-0.47$,  then the iron mass fraction in it 
$Z_{Fe}=10^{-0.47-2.84}=10^{-3.31}$.
In other words, the iron abundance in the metal-rich GC group is fifteen times higher 
than that in the metal-poor. Now we assume that the iron mass fraction in the 
clouds is the same as in GCs. If the mass of a cloud $M\approx10^9 M_{\sun}$, 
then the iron mass $M_{Fe}\approx10^{4.5} M_{\sun}$ in the metal-poor cloud.

Let us, first, consider the case when the fraction of the enriched part of the cloud is
20\% of the initial cloud mass (first possibility formulated in Sec.~\ref{sec_4_1}).
Then the iron mass has increased from $M_{Fe}=10^{8.3-4.5}=10^{3.8} M_{\sun}$
to $M_{Fe}=10^{8.3-3.31}\approx10^{5} M_{\sun}$, and the cloud gained 
$M_{Fe}=10^{4.97}\approx10^{5} M_{\sun}$.

 When a core-collapse supernova explodes, $M_{Fe}\approx0.01 M_{\sun}$ is emitted \citep{t95}. Hence, $10^{5.8}$ \mbox{SNe~CC} will produce 
$M_{Fe}=10^{3.8} M_{\sun}$. Thus, the contribution of core-collapse supernovae to the production of iron is negligibly small.
 Consequently,  $M_{Fe}\approx10^{5}  M_{\sun}$ should be produced by other sources -- \mbox{SNe~Ia}, 
 and mainly by those formed from short-lived 
progenitors -- prompt \mbox{SNe~Ia} (\mbox{pSNe~Ia}), since according to \citet{m06}, \citet{l11}, \citet{m14}, \citet{r14}, and \citet{ah18},
\mbox{SNe~Ia} evolved from long-lived progenitors have not been observed for initial 1-1.5 Gyr. 
If we assume that the canonically accepted $0.6 M_{\sun}$ of iron is emitted with 
the explosion of such a \mbox{pSNe~Ia} \citep{n97}, then
$10^{5.2}$ of \mbox{pSNe~Ia} will be necessary which is four times smaller than \mbox{SNe~CC}. 
If we assume that with the \mbox{pSNe~Ia} explosion, $0.23 M_{\sun}$ 
of iron is emitted (it is the mass gained according to theoretical calculations conducted in \citet{a13} 
for \mbox{pSNe~Ia} and complying with the estimate of 
the nickel mass emitted in explosions of some \mbox{SNe~Ia} \citep{c15}), then $10^{5.6}$ of \mbox{pSNe~Ia} will be needed 
which is  1.6 times smaller than the number of \mbox{SNe~CC}. 
For comparison, according to \citet{l11}, the current frequency of explosions of \mbox{SNe~CC} in our 
Galaxy equals 2.3 per century, and the frequency of bursts of \mbox{pSNe~Ia} in star-forming regions is 0.43 per 
century. In other words,  in massive galaxies, the frequency of bursts of \mbox{SNe~CC} is five times higher than 
those of \mbox{pSNe~Ia}.
It is important to note that the conclusions on the frequency of supernovae explosions 
in \citet{l11} were drawn from the analysis of evolved galaxies close to the modern age. 
We, however, are concerned with the GCs, in which the relation between the supernovae subtypes 
is not measured observationally.
The characteristic property of GCs is in the fact that the density of stars is high in them.
When analysing the obtained results, we conclude, that our first assumption leads 
to the reliable estimates of the frequencies of \mbox{SNe~CC} and \mbox{pSNe~Ia} bursts and could take place in reality.

 We now turn to the second possibility, formulated in Sec.~\ref{sec_4_1}: the
mass of the enriched part of the cloud is twice less than the original one
and fraction of the cloud gas transformed into stars should be $\sim 25\%$.
In this case, the mass of iron in the enriched part of the cloud is $M_{Fe}=10^{5.4} M_{\sun}$. 
It can be accepted that the cloud gained $M_{Fe}=10^{5.37} M_{\sun}$. 
The contribution to the iron production by $10^{6.2}$~\mbox{SNe~CC} is negligible, as it was argued above. 
Consequently, when producing $0.6 M_{\sun}$ of iron by a single supernova, the necessary number of \mbox{pSNe~Ia} is  $10^{5.6}$ 
which is four times smaller than the number of \mbox{SNe~CC}.  On the other hand,  
when producing $0.23 M_{\sun}$ of iron by a single supernova, 
the necessary number of \mbox{pSNe~Ia} is $10^6$, which is $1.6$ times less than the number of \mbox{SNe~CC}. 
According to the presented analysis of the frequencies of \mbox{SNe~CC} and \mbox{pSNe~Ia} explosions, 
the ratio of frequencies has not changed with respect to the first considered possibility.
One may conclude that the second cloud enrichment scenario is probable.

From the above arguments, it follows automatically that the third possibility 
formulated in Sec.~\ref{sec_4_1} is impossible, because
a number of \mbox{pSNe~Ia} cannot six times exceed a number of \mbox{SNe~CC}. 

Can restrictions be placed on the frequency of different types of supernovae 
from analysing of other properties of globular clusters? 
It is the characteristic feature of GCs that the density of stars is high 
in them and stellar collisions are possible \citep{hd76, f04}. 
Some researchers consider blue straggler stars to be an observational manifestation 
of such processes \citep{f09, d13, s14, l18}. 
It is important to notice that the exact nature of stars that produce \mbox{SNe~Ia}, 
remains unknown \citep{r11, hn00, l00}. It is generally agreed that 
type \mbox{SNe~Ia} is caused by an explosion of a carbon-oxygen white dwarf. However, 
the details of the process of accretion of companions matter onto it remain unclear 
\citep{y08}. At the same time, the recent finding by \citet{s12} shows that the 
binary interaction may affect 70\% or more of the massive star population. 
This result is confirmed by the studies of \citet{kf07}. At the same time, 
the interaction of binary stars is considered to be an important formation channel
of SN Ibc supernovae also \citep{n95}, whose spectra show the high magnesium abundance 
and the low iron abundance \citep{s09}. 
Thus, it can be concluded that the available data, unfortunately, 
are not enough to predict the ratio of the \mbox{SNe~CC} and \mbox{pSNe~Ia} frequencies. 
Analysis of abundances of chemical elements is the only source of information 
about the number of different types of supernovae in GCs.

\section[]{Conclusions}

We have conducted a statistical study of the iron and magnesium abundances in GCs 
circumgalactic gaseous clouds. 
On the comparison between the magnesium abundance in the metal-poor and metal-rich subgroups 
of circumstellar clouds at redshifts of $0.1<z<1.1$ 
with magnesium and the magnesium abundance in similar subgroups of GCs, we can conclude that they
coincide within an error of abundance measurements 
in clouds of 0.3 dex. The magnesium abundance in clouds is by 0.14 dex smaller than in GCs 
for the metal-rich subgroup and by 0.18 dex -- 
for the metal-poor subgroup. An assumption can be made from this that the pLLSs+LLSs 
clouds are the remnants of host clouds of GCs.

From the fact that the cloud metallicity distribution is presented only by the metal-poor 
component at redshifts $2.5<z<3.3$, it follows that the 
fraction of metal-poor gas was enriched with heavier elements through thermonuclear 
fusion products of supernovae of the first GC generation entering into it. 
The second generation of GCs was formed from this enriched gas.

Let us discuss how the scheme proposed in our study agrees with ideas developed in the literature.
In the literature, the process of formation of globular clusters is considered within 
the framework of two basic concepts. 
According to the first one, metal-poor GCs are formed in low-mass galaxies and then, 
together with parent galaxies, are accreted to the halo of massive galaxies
(\cite{sz78}, \cite{s10} and references therein). 
Metal-rich GCs form from the gas enriched by stars of massive galaxies. These stars were not constituents
of metal-poor GCs (see also \cite{el12}).
According to the second concept, the halo of each massive galaxy 
experiences two episodes of the GC formation (the `in situ' model) \citep{fbg97}. 
First, when a proto-galactic cloud collapses, low-metallicity GCs are formed. 
High-metallicity GCs originate during the second stage of star formation.

The results of our study are consistent with the second aforementioned concept \citep{fbg97}.

Our reasoning can be relevant not only for GCs but also in the case of 
ultra-compact dwarf galaxies (UCDs),
because the last ones have metallicities, spatial distribution and specific frequencies 
similar to those of GCs 
(e.g. \cite{Zhang18}, \cite{Voggel16}, \cite{Mieske12}, \cite{Francis12} 
and references in these papers).
UCDs are more massive, luminous, and have higher mass-to-light ratios than globular clusters
but are fainter and more compact than dwarf elliptical galaxies and M32-like galaxies.
It was proposed many times in the literature that UCDs and GCs 
were formed under similar conditions
(\cite{gb18}, \cite{Romanowsky17}, \cite{Goerdt08} \cite{c12} and references in these papers).

Unfortunately, the available literature data about the properties of pLLSs 
are not sufficient to make definite conclusions
concerning the spatial distribution of their low-metallicity and high-metallicity 
representatives relative to the planes of galactic disks.
It is clear that DLAs ($logN_{HI}>20.3$) are more concentrated to the planes of galactic disks.
The typical distances from host galactic planes for DLAs are comparable with these of thick disk objects 
in our Galaxy \citep{bel12}.
DLAs are on average denser than pLLSs. The mean metallicity of DLAs is close to the metallicity of high-metallicity pLLSs.
We cannot extrapolate our conclusions to DLAs, because there are no enough data 
in the literature for the statistical analysis of magnesium abundances in them (e.g. \cite{Vladilo11}). 

Studying the abundances of chemical elements in GCs, one can come to nontrivial conclusions 
regarding the fraction of gas transformed into stars, the contribution of supernovae 
to the production of heavier elements and
the fraction of the circumgalactic gas enriched in chemical elements synthesized in it during first starforming bursts. 
The amount of magnesium produced by first generations of GCs does agree with two hypotheses: 
i) the fraction of mass of the enriched part of the cloud is $20\%$ of the initial cloud mass
and the fraction of the cloud gas, transformed into stars is $10\%$;
ii) the mass of the enriched part of the cloud is twice less than the original one, 
and the fraction of the cloud gas transformed into stars is $25\%$.

According to the available observations and theoretical calculations, 
\mbox{SNe~CC} produce a negligible amount of iron. The bulk of its mass is produced by \mbox{SNe~Ia}, and namely
by those the progenitors of which explode in the first Myr after forming. 
It is shown in the present study that the number of \mbox{pSNe~Ia} in GCs must be
2-4 times less than the number of \mbox{SNe~CC}.

\section*{Acknowledgments}
 We thank the anonymous referee for valuable comments that helped to improve the paper.
The work is supported by the Russian Foundation for Basic Research (grant no. RFBR 18-02-00167). 
A. I. A. is thankful to E. O. Vasiliev (SFedU) for useful discussions about mechanisms of metal mixing in 
circumgalactic clouds.

\bsp

\label{lastpage}

\end{document}